\documentclass[aps,prl,english,twoside,twocolumn]{revtex4}
\usepackage[T1]{fontenc}
\usepackage[latin1]{inputenc}
\usepackage{babel}
\usepackage{epsfig}
\usepackage{graphicx}
\usepackage{slashbox}
\makeatletter

\providecommand{\tabularnewline}{\\}

\newcommand{\ket}[1]{| #1 \rangle}
\newcommand{\bra}[1]{\langle #1 |}

\newcommand{\be}{\begin{equation}} \newcommand{\ee}{\end{equation}}
\newcommand{\ba}{\begin{eqnarray}} \newcommand{\ea}{\end{eqnarray}}

\newcommand{\etal}{\emph{~et~al.}}
 \def\upa{\uparrow}
 \def\doa{\downarrow}

\def\be{\begin{equation}}
\def\ee{\end{equation}} 
\def\beq{\begin{equation}}
\def\eeq{\end{equation}}
\def\bea{\begin{eqnarray}} 
\def\eea{\end{eqnarray}}

\def\F{\mathcal{F}}

\def\[{\{}
 
\def\]{\}}

\makeatother
\begin{document}

\title{Measurement-Based Teleportation Along Quantum Spin Chains}

\author{J.P. Barjaktarevic}
\email{jpb@physics.uq.edu.au}
\affiliation{Department of Physics, University of Queensland, Brisbane, Queensland QLD 4072, Australia}
\author{J. Links}  
\affiliation{Department of Mathematics, University of Queensland, Brisbane, Queensland QLD 4072, Australia}
\author{G.J. Milburn}  
\affiliation{Center for Quantum Computer Technology, School of Physical Sciences, The University of Queensland, QLD 4072 Australia}
\author{R. H. McKenzie}
\affiliation{Department of Physics, University of Queensland, Brisbane, Queensland QLD 4072, Australia}

\date{\today{}}

\begin{abstract}

We consider teleportation of an arbitrary spin-$\frac{1}{2}$ target quantum state along the ground state of
a quantum spin chain. We present a decomposition of the Hilbert space of the many-body quantum states into
$4$ vector spaces.  Within each of these subspaces it is possible to take any superposition of states, and
use projective measurements to perform unit fidelity teleportation.  We also show that all total spin-$0$ many-body states belong in the same space, so it
is possible to perform unit fidelity teleportation over any one-dimensional spin-$0$ many-body 
state.  We generalise to $n$-Bell states, and present some general bounds on fidelity of teleportation given
a general state of a quantum spin chain.

\end{abstract}
\maketitle

Quantum many-body spin Hamiltonians often have highly entangled ground states, and an
understanding of entanglement may in turn lead to a greater understanding of the physics of many-body
quantum systems with strong correlations\cite{cmp_ent}. However, a definitive measure of multipartite
entanglement is yet to be found, though there exist several proposed measures \cite{
multipartite_ent_measure1, multipartite_ent_measure2, multipartite_ent_measure3}.  In this Letter, we
characterise the entanglement content of the ground states of several quantum spin chain Hamiltonians by
its ability to teleport a state to an arbitrary site.

The protocol used for teleportation will be based purely upon measurement, local unitary operations, and
classical communication.  It has recently been noted that these operations are sufficient
for universal QC (quantum computation)\cite{nielsen_proj_meas, raussendorf_briegel, general_proj_meas},
particularly as manifest in cluster state QC.  Hence, characterising the ability of a system to perform high
fidelity teleportation is useful for the purposes of QC.
 
Teleportation over any Bell state by Bell basis measurements is well understood\cite{teleportation}. We begin
extending this protocol by considering $\mathcal{L}$ pairs of particles in tensor products of Bell states, which provide a basis for the
Hilbert space.  We then show that we can decompose this space into classes of states which are equivalent for the purposes
of teleportation. The ground states of several quantum spin chain Hamiltonians are considered.  Finally, we introduce a two component quantity, whose magnitude lower
bounds the average fidelity of a teleported state.

\emph{Tensored Bell basis teleportation}. Let us denote our $2$-dimensional spin space as $V$, with basis
$\left|\upa\right>,\,\left|\doa\right>$.  Let $v^i$ represent the standard Bell basis states, where the $v^0$ is the
singlet state,

\beq v^0=\frac{1}{\sqrt{2}}\left(\left|\upa\right>\left|\doa\right>
-\left|\doa\right>\left|\upa\right>\right) . \label{singlet}\eeq

The other Bell states, $v^i$ may be written as $v^i=\left(I\otimes X^i\right)v^0$, where $X^0=\sigma_I$,
$X^1=\sigma_x$, $X^2=-\sigma_z$, and $X^3=i \sigma_y$, where
the $\sigma$ matrices are the usual Pauli matrices
for spin-$\frac{1}{2}$ particles.
These states are simultaneous eigenstates of $\sigma^x \otimes \sigma^x$ and $\sigma^z \otimes \sigma^z$.
Let us introduce
the notation $\left|kl\right\}$ for Bell states such
that $\sigma^x\otimes \sigma^x \left|kl\right\}=k \left|kl\right\}$,
and $\sigma^z\otimes \sigma^z \left|kl\right\}=l \left|kl\right\}$. Multipartite
states which are
amenable to use for teleportation turn out to be simultaneous eigenstates of similar $\mathcal{L}$-qubit operators.

\begin{figure}[t]
  \begin{center}
    \includegraphics[width=2.0in]{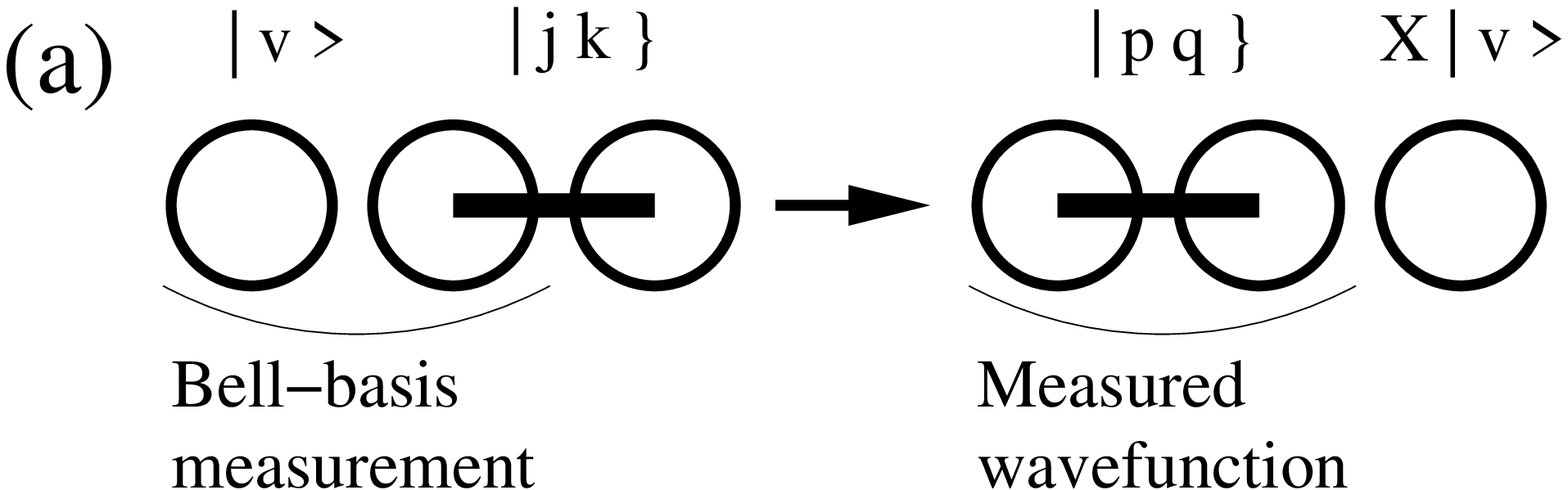}

    \includegraphics[width=3.0in]{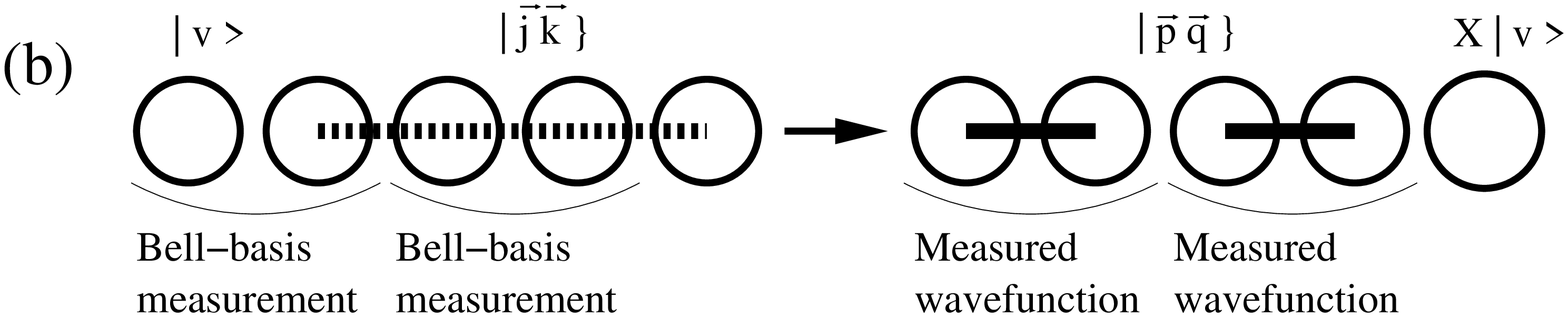}
  \end{center}
\caption{A schematic representation of the proposed teleportation protocol, showing input state $\ket{v}$ and final state $X 
\ket{v}$.  $X$ is known from the measurement outcome, (see Table \ref{corr_table}).  (a) A Bell basis measurement over a 
single pair of maximally entangled spins, as in Eq.
\ref{single_measurement}, represented by a solid line between two sites. (b) A schematic representation of our protocol
for a general spin quantum state, as in Eq. \ref{many_measurements}, with
non-trivial entanglement, represented by the dashed line.  Here we assume that the state
$\left|\vec{j}\vec{k}\right\}$ lies within a Bell subspace, yielding a pure output state, $X \ket{v}$.}

  \label{fig1}
\end{figure}

Let $P^{\left|{jk}\right\}}$ denote the projections onto 
the state $\left|jk\right\}$.  Given a target state which we wish to teleport, $\left|v\right>$, and a two particle Bell state $\left|jk\right\}$ over
which we wish to teleport it, we can simply perform a Bell-basis measurement across the target state, and half of
the two-particle state, as in Fig. \ref{fig1} (a).  We can make the following decomposition:

\beq \left|v\right> \otimes \left|jk\right\}= \frac{1}{2}\sum_{p,q} \left|pq\right\}\otimes X^{jk}_{pq}\left|v\right>. \label{single_measurement} \eeq
where $X^{jk}_{pq}$ is one of the matrices $X^i$, as in Table \ref{corr_table}.  We can identify 
$X^{jk}_{pq}$ with a unitary correction, which we must invert to yield the target state
$\left|v\right>$.  We now extend this 
standard result\cite{teleportation} to
larger systems by making use of the fact that the matrices $X^i$ form a group under composition.

\begin{table}
\begin{tabular}{|c|c|c|c|c|}
\hline
\backslashbox{ $P^{\left|pq\right\}}$ }{ $\left|jk\right\}$ }  & $\left|--\right\}$ & $\left|-+\right\}$ & $\left|+-\right\}$ & $\left|++\right\}$\tabularnewline \hline
$P^{\left|--\right\}}$ & $-X^0$ & $-X^1$ & $-X^2$ & $-X^3$\tabularnewline
\hline
$P^{\left|-+\right\}}$ & $X^1$ & $X^0$ & $-X^3$ & $-X^2$\tabularnewline
\hline
$P^{\left|+-\right\}}$ & $X^2$ & $X^3$ & $X^0$ & $X^1$\tabularnewline
\hline
$P^{\left|++\right\}}$ & $-X^3$ & $-X^2$ & $X^1$ & $X^0$\tabularnewline
\hline
\end{tabular}

\caption{Unitary correction, $X^{jk}_{pq}$, to be inverted on the final qubit, where we start with a quantum state $
\left|v\right> \otimes \left|jk\right\}$ and project onto $\left|pq\right\}$ with $P^{\left|pq\right\}} = \left|pq\right\}\left\{pq\right| \otimes I$.  These
correspond to either no correction ($X^0$), a bit flip ($X^1$), a phase flip ($X^2$), or both a bit and a phase flip ($X^3$)}

\label{corr_table}
\end{table}

For an even number $L=2\mathcal{L}$ of spin-$\frac{1}{2}$ particles, we define the
$L$-particle state $\left|\vec{j}\vec{k}\right\} = \left|j_1 k_1\right\}\otimes \dots \otimes \left|j_\mathcal{L} k_\mathcal{L}\right\}$, where
$\vec{j} = (j_1,j_2, \dots ,j_{\mathcal{L}})$ is a
$\mathcal{L}$-dimensional vector. We
can then make the following decomposition

\beq
\left|v\right>\otimes \left|\right.\vec{j} \vec{k} \left.\right\}
=\frac{1}{2^{\mathcal{L}}}\sum_{\vec{p},\vec{q}}
\left|\right.\vec{p}\vec{q}\left.\right\}\otimes X^{j_\mathcal{L} k_L}_{p_\mathcal{L} q_\mathcal{L}}\dots
X^{j_1k_1}_{p_1q_1} \left|v\right> \label{many_measurements} \eeq
Since the corrections $X^i$ are closed under composition, we may write $ X^{j_\mathcal{L} k_\mathcal{L}}_{p_\mathcal{L} q_\mathcal{L}}\dots
X^{j_1k_1}_{p_1q_1} = X^{\vec{j}\vec{k}}_{\vec{p}\vec{q}} = \pm X^i$ for some $i$.  We may decompose the $L$-qubit Hilbert
space $V^L$ into $4$ subspaces, each of which has a different total correction $X^{i}$.  For any state within a
given subspace $V^L_{[kl]}$, we are able to perform unit fidelity teleportation with any set of measurement results.  This decomposition turns out
to be independent of $\vec{p}$ and $\vec{q}$
\footnote{For any state $\left|jk\right>$, we may change the projection from $P^{pq}$ to $P^{rs}$, and the total unitary
correction will change by $X^{jk}_{pq} (X^{jk}_{rs})^{-1}$, which is independent of $\{j,k\}$.}.  Decomposing the Hilbert space
as

\beq V^L= V^L_{[--]}\oplus V^L_{[-+]}\oplus V^L_{[+-]} \oplus V^L_{[++]}, \label{decomposition} \eeq
the corrections are in $1$-$1$ 
correspondence with the assignments given in Table \ref{corr_table}.

Physically, each component of the decomposition corresponds to all of the states which will give the same 
unitary correction after the
series of Bell basis measurements.  We
refer to the 4 subspaces as Bell subspaces since they reduce to the Bell
basis in the two-qubit case.

The decomposition of $V^L$ into $4$ subspaces with different corrections turns out to be extremely useful in understanding
systems which have ground states lying entirely in one of these subspaces.  We present several model Hamiltonians for
which this is true.  It is interesting to note that all of these ground 
states are similar to spin liquid states\cite{spin_liquid}, since every member of
$V^L_{[kl]}$ has every $1$-qubit reduced density matrix equal to $\frac{1}{2} I$.  Hence each site is also maximally
entangled with the rest of the chain.  Any state in one of these subspaces has maximum localisable 
entanglement with respect to Bell measurements\cite{localizable_entanglement}.

\emph{Spin-$\frac{1}{2}$ Next-Neighbour Hamiltonians}. A family of nearest- and next-nearest neighbour 
antiferromagnetic spin 
exchange
Hamiltonians is parameterised, with $\beta>0$ by
\beq H_{n,nn} = \sum_{i=1}^{N}\hat{S}_{i}.\hat{S}_{i+1}+\beta\hat{S}_{i}.\hat{S}_{i+2} \label{mg_family} \eeq
for which $\beta=\frac{1}{2}$ yields the Majumdar-Ghosh Hamiltonian.  
The ground state of the Majumdar-Ghosh Hamiltonian\cite{mg} is simply comprised of a tensor product of singlets,
$\otimes_{k=1}^{N/2}\ket{v^0} $, and one may use this to perform unit fidelity teleportation with repeated Bell basis
measurements along the chain.  However, it is possible to show that the ground state of Eq.  \ref{mg_family} for any value
of $\beta>0$ lies within a Bell subspace, including the specific case of $\beta=0$, which corresponds to the Heisenberg
Hamiltonian.

More generally, it can be shown rigorously that any quantum state with a total spin-$0$ lies within one of the
Bell subspaces, $V^L_{[kl]}$, which we outline as follows.  Given $\Psi^0=v^0\otimes \dots
\otimes v^0$, it is clear that $\Psi^0$ must have total spin $0$.  Further, any permutation of
sites in $\Psi^0$ must also have spin $0$, and belong to the same subspace, $V^L_{[j^0 k^0]}$\footnote{More
formally the spaces, $V^L_{[jk]}$, are invariant under an action of the symmetric group}.  In fact, it is
possible to decompose every spin-$0$ state as a sum over permutations of the sites of $\Psi^0$\cite{rvb_singlet}. 
This is known in the chemistry literature as resonant valence bond theory\cite{soos_ramasesha}.


It is interesting to note that for the antiferromagnetic Ising Hamiltonian\cite{Ising}, there are two degenerate product
form ground states, $\left|\Phi^-\right>=\left|\upa\doa\dots\doa\right>$ and $\left|\Phi^+\right>=\left|\doa\upa\dots\upa\right>$.  The decomposition of
these will have equal support in more than one Bell subspace.

$$
\left|\Phi^\pm\right>=(\left|+-\right\}\pm\left|--\right\})\otimes\dots\otimes(\left|+-\right\}\pm\left|--\right\})$$
However, the superposition, $\ket{\Phi^0}=\frac{1}{\sqrt{2}}(\left|\Phi^+\right>-\left|\Phi^-\right>)$ has support on only 
one Bell subspace.  Further, $\ket{\Phi^0}$ has only one ebit\footnote{1 ebit is equivalent to one distillable singlet.} of
entanglement - yet even so, we can use this state to perform unit fidelity teleportation over an arbitrary distance.  

\emph{Spin-$1$ Models}. The AKLT model\cite{aklt} can be related to the antiferromagnetic Heisenberg model 
for spin-$1$, and 
is given by
the specific case of $\alpha=\frac{1}{3}$ in the class of Hamiltonians 

\beq H_{AKLT} = \sum_{i=1}^{N}\hat{S}_{i}.\hat{S}_{i+1}+\alpha(\hat{S}_{i}.\hat{S}_{i+1})^{2} \label{AKLT} \eeq

We may decompose each spin-$1$ site, $i$, as two virtual spin-$\frac{1}{2}$ sites, $i,\bar{i}$, and project onto the spin-$1$
subspace.  For an $N$ site chain with spin-$\frac{1}{2}$ boundary conditions, we may write the ground state of the AKLT model
as\cite{div_ent_length}

\beq \ket{\psi_{AKLT}} = (\otimes_{k=1}^N A_{k \bar{k}}) \ket{I} \eeq
where $\ket{I} = \otimes_{k=0}^N \ket{I_{\bar{k} k+1} }$ is a product of singlets, and $A_{k \bar{k}}$ projects the
spins at sites $k$ and $\bar{k}$ onto their symmetric subspace.  The operation of projecting out the singlet components
leaves the state in the Bell subspace which also contains the singlet quantum states.

Verstraete \etal\cite{div_ent_length} show that the projection onto Bell states in the $i \otimes \bar{i}$ space is achievable
with only single particle measurements.  Hence, one can use only single particle spin-$1$ measurement to 
teleport a
spin-$\frac{1}{2}$ state along a AKLT chain.  Moreover, a linear spin-$1$ Heisenberg antiferromagnetic may exhibit
spin-$\frac{1}{2}$ degrees of freedom at the boundaries, as evidenced by both numerical\cite{dmrg} and 
experimental
results\cite{spinhalf_ends}.  It may be possible to couple the target state into the spin chain by a Bell basis
measurement over the target state and the boundary spin-$\frac{1}{2}$ degrees of freedom.  Hence the entire teleportation problem may
reduce to an initial Bell basis measurement and single particle measurements on spin-$1$ particles (see also \cite{fan_aklt}).

\emph{Non-Bell basis Measurements}. Experimentally, it is very difficult to perform Bell basis measurements 
directly, and we now consider
teleportation using only single particle measurements.  To affect a Bell basis measurement, we may use the fact
that there is a similarity transformation between Bell basis projection 
operators and a product of single particle
projections, by using an entangling operation, $U$:

\beq P^j = U^\dagger P^{j_2}_2 P^{j_1}_1 U \eeq
where $j_1$ and $j_2$ are functions of $j$ set by the unitary, $U$.  Projections onto different sites 
commute, and hence, we may decompose our Bell measurement protocol as an
application of entangling unitaries onto the whole quantum state, followed by a complete set of single particle
projective measurements.  The reduced
requirements of single particle measurements and entangling operations are very similar to the requirements for
cluster state teleportation\cite{raussendorf_briegel}.

\emph{Fidelity}. Our protocol involves using a channel quantum state, $\ket{\psi}$, which we believe to 
belong to the
subspace $V^L_{[pq]}$, performing Bell basis measurements along the chain, and then applying
the appropriate correction.  Let us decompose our channel state as

\beq\ket{\psi}=\sum_{i,j=\pm} c_{i,j} \ket{\phi_{i,j}} \label{almost_perfect}\eeq
where $\ket{\phi_{i,j}}$ are orthonormal
vectors, each lying in subspaces $V^L_{[ij]}$.  The condition that $\ket{\psi}$ lies within the
subspace $V^L_{[pq]}$ implies $c_{i,j} = \delta_{i,p} \delta_{j,q}$, which is only approximately satisfied for real 
quantum states.  Projecting into the Bell basis yields the measurement result $\left|\vec{a}\vec{b}\right\}$ with 
probability, $p_{
\left|\vec{a}\vec{b}\right\} }$.  If the target quantum state is $\left|v\right>$, then the resulting quantum 
state will
be $\left| \vec{a}\vec{b} \right\} \otimes \ket{v'}$ where $\ket{v'_{\vec{a} \vec{b}}} \propto \sum_{i,j=\pm} c_{i,j}
X_{\vec{a}\vec{b}}^{ij} \ket{v} $ is a superposition of different corrections onto 
$\left|v\right>$.  The fidelity for a particular state $\ket{v}$
is given by $\F=\left| \bra{v'_{\vec{a} 
\vec{b}}}X^{pq}_{ab}\ket{v} \right|^2 $. 

\emph{Teleportation Parameter}. It is desirable to a define a quantity which we can use to
characterise the fidelity of teleportation achievable by a measurement protocol over a quantum state.  For 
this purpose, for any quantum spin state $\ket{\Psi_0}$, we define a quantity

\beq \mathcal{O}(\Psi_0) = \sum_{\alpha=x,y,z}\left|\bra{\Psi_0}\otimes_{k=1}^L \sigma^\alpha_k \ket{\Psi_0} \right |
\label{odef}\eeq
where $\left| \mathcal{O} \right| = 4 \left|c_{p,q}\right|^2-1$.

We note that $\mathcal{O}(\Psi_0)$ is related to non-local correlations in $\ket{\Psi_0}$, bearing some similarity to 
a string order parameter\cite{preroughening}.  We now show that $\mathcal{O}(\Psi_0)$ can be used to give a lower 
bound on the 
fidelity of teleportation through $\ket{\Psi_0}$.
When $\mathcal{O}$ is close to $3$, we satisfy the condition $c_{i,j} \simeq \delta_{i,p} \delta_{j,q}$, 
with equality when
$\mathcal{O}=3$.

\begin{figure}[t]
  \begin{center}
    \includegraphics[width=3.0in]{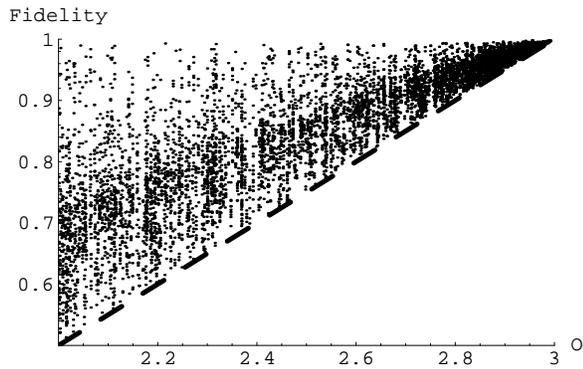}
  \end{center}
  \caption{Data points correspond to the fidelity of teleportation for a range of target states.  The fidelity 
is shown against the quantity $\mathcal{O} $ as defined in Eq. \ref{odef} for 2000 $4$-qubit channels.  The dashed line is a lower bound for these points and is given by the inequality in Eq. 
\ref{f_bound}.
This clearly shows a tendency towards a fidelity, $\F=1$ as $\mathcal{O}$ tends to $3$.  
}
  \label{fig2}
\end{figure}

The minimum possible fidelity given perfect Bell basis projections is a monotonically decreasing function 
of $ \mathcal{O} $, as seen in Fig. 
\ref{fig2}.  Note that
the fidelity $\F$ is not uniquely defined by $  \mathcal{O} $, also being dependent on $\ket{v}$ and the measurement history.  Furthermore, for an arbitrary channel, 
measurement in the Bell basis is not
necessarily optimal.  These facts are reflected by the following inequality:

\beq
\F \ge \frac{\mathcal{O}-1}{2}
\label{f_bound}
\eeq

\emph{Generalisations}. Generalisation to higher spin is possible, where a generalised $N$-Bell state is 
given by $$ \left|
A B \right\} = \sum_{i=0}^{N-1} \omega^{B i} \ket{i(i+A)} \label{nbell} $$ where $s=\frac{N-1}{2}$ is the spin,
$\omega=\sqrt[N]{1}$ is the primitive $N$th root of $1$, and $A, B \in \{-s,\dots,s\}$ and $\ket{ij}$ are the usual
two-particle spin-$s$ basis states.
In analogy to the interpretation of $\sigma_x$ as a bit flip, and and $\sigma_z$ as a phase flip, we
introduce the permutation and phase correction matrices as:

$$P=\pmatrix{
0 & 1 & 0 & \dots & 0 \cr
0 & 0 & 1 & \dots & 0 \cr
\vdots & \vdots & \vdots & \ddots & \vdots \cr
0 & 0 & 0 & \dots & 1 \cr
1 & 0 & 0 & \dots & 0
}
$$

$$Q=\pmatrix{
1 & 0 & \dots & 0 \cr
0 & \omega & \dots & 0 \cr
\vdots & \vdots & \ddots & \vdots \cr
0 & 0 & \dots & \omega^{N-1}
}
$$
As in the case of spin-$\frac{1}{2}$ particles, the corrections $P^i$ and $Q^i$ for $i=1,2,\dots,N$ form a 
group, and we may perform exactly the same procedure of
generalised Bell-basis measurement, followed by a cumulative correction at the end of the procedure.

\emph{Cluster State QC}. By inspection, the regular $2$-site cluster state \cite{raussendorf_briegel},
$\ket{\doa \doa} + \ket{\doa \upa} + \ket{\upa \doa} - \ket{\upa \upa}$ does not lie in any of the Bell 
subspaces,
$V^L_{[i,j]}$, since it is a linear superposition of two Bell basis states.  However, we may rewrite this state as
$\ket{\doa \rightarrow} + \ket{\upa \leftarrow}$, where $\ket{\rightarrow}=\ket{\doa}+\ket{\upa}$ and
$\ket{\leftarrow}=\ket{\doa}-\ket{\upa}$ are the usual eigenvectors of $\sigma_x$.  This state now resembles a Bell
state, and we may repeat our entire construction by using the $z,x$ basis rather than the $z,z$ basis.  More generally 
it can be shown that the $1$-dimensional $L$-qubit cluster states of \cite{raussendorf_briegel} can be mapped to the 
subspace $V^L_{[++]}$ through a sequence of local unitary transformations.  This may be expected, as our main result 
strongly resembles cluster state methods in the use of local measurement and feed forward.  The main difference is in 
the initialisation of the cluster state, in which entanglement is provided by a series of controlled-phase entangling 
gates, in contrast to our requirement that the spin system is in a Bell subspace.

We have shown that by using a Bell-basis measurement protocol, it is possible to 
decompose
the Hilbert space into Bell subspaces, using a decomposition into independent corrections. Further, 
we have
shown that if any quantum state belongs to a Bell subspace, then we may perform unit fidelity teleportation 
with it,
using only measurements in the generalised Bell basis.  We have presented several model Hamiltonians for which the
ground state is amenable to our teleportation protocol.  Some alternative measurement schemes have been presented,
including a scheme for a spin-$1$ system.  Finally, we have presented a 
complex number valued parameter whose magnitude provides a lower bound on the fidelity of teleportation using our 
measurement protocol.  Finally, we note that we have only considered the ability to teleport one spin 
quantum state to an arbitrary
site\footnote{We have not assumed any special ordering on the sites or the measurements.} within a spin chain.  Thus, $v^0$ and $v^0
\otimes v^0$ both have $\mathcal{O} = 3$, despite having a different number of ebits.  There may 
be further insights to be gained
into the entanglement content of a system, by considering the ability to teleport several spin
states to arbitrary sites.

This work
was supported by the Australian Research Council.  JPB thanks G. A. D. Briggs and the QIPIRC at the University of Oxford for very kind hospitality.


\begin{thebibliography}{21}
\expandafter\ifx\csname natexlab\endcsname\relax\def\natexlab#1{#1}\fi
\expandafter\ifx\csname bibnamefont\endcsname\relax
  \def\bibnamefont#1{#1}\fi
\expandafter\ifx\csname bibfnamefont\endcsname\relax
  \def\bibfnamefont#1{#1}\fi
\expandafter\ifx\csname citenamefont\endcsname\relax
  \def\citenamefont#1{#1}\fi
\expandafter\ifx\csname url\endcsname\relax
  \def\url#1{\texttt{#1}}\fi
\expandafter\ifx\csname urlprefix\endcsname\relax\def\urlprefix{URL }\fi
\providecommand{\bibinfo}[2]{#2}
\providecommand{\eprint}[2][]{\url{#2}}

\bibitem[{\citenamefont{Preskill}(2000)}]{cmp_ent}
\bibinfo{author}{\bibfnamefont{J.}~\bibnamefont{Preskill}},
  \bibinfo{journal}{J. Mod. Optics} \textbf{\bibinfo{volume}{47}},
  \bibinfo{pages}{127} (\bibinfo{year}{2000}).

\bibitem[{\citenamefont{Vedral et~al.}(1997)\citenamefont{Vedral, Plenio,
  Rippen, and Knight}}]{multipartite_ent_measure1}
\bibinfo{author}{\bibfnamefont{V.}~\bibnamefont{Vedral}},
  \bibinfo{author}{\bibfnamefont{M.~B.} \bibnamefont{Plenio}},
  \bibinfo{author}{\bibfnamefont{M.~A.} \bibnamefont{Rippen}},
  \bibnamefont{and} \bibinfo{author}{\bibfnamefont{P.~L.}
  \bibnamefont{Knight}}, \bibinfo{journal}{Phys. Rev. Lett.}
  \textbf{\bibinfo{volume}{78}}, \bibinfo{pages}{2275} (\bibinfo{year}{1997}).

\bibitem[{\citenamefont{Wong and
  Christensen}(2001)}]{multipartite_ent_measure2}
\bibinfo{author}{\bibfnamefont{A.}~\bibnamefont{Wong}} \bibnamefont{and}
  \bibinfo{author}{\bibfnamefont{N.}~\bibnamefont{Christensen}},
  \bibinfo{journal}{Phys. Rev. A} \textbf{\bibinfo{volume}{63}},
  \bibinfo{pages}{044301} (\bibinfo{year}{2001}).

\bibitem[{\citenamefont{Biham et~al.}(2002)\citenamefont{Biham, Nielsen, and
  Osborne}}]{multipartite_ent_measure3}
\bibinfo{author}{\bibfnamefont{O.}~\bibnamefont{Biham}},
  \bibinfo{author}{\bibfnamefont{M.~A.} \bibnamefont{Nielsen}},
  \bibnamefont{and} \bibinfo{author}{\bibfnamefont{T.~J.}
  \bibnamefont{Osborne}}, \bibinfo{journal}{Phys. Rev. A}
  \textbf{\bibinfo{volume}{65}}, \bibinfo{pages}{062312}
  (\bibinfo{year}{2002}).

\bibitem[{\citenamefont{Nielsen}(2003)}]{nielsen_proj_meas}
\bibinfo{author}{\bibfnamefont{M.~A.} \bibnamefont{Nielsen}},
  \bibinfo{journal}{Phys. Lett. A} \textbf{\bibinfo{volume}{308}}
  (\bibinfo{year}{2003}), \eprint{quant-ph/0108020}.

\bibitem[{\citenamefont{Raussendorf and Briegel}(2001)}]{raussendorf_briegel}
\bibinfo{author}{\bibfnamefont{R.}~\bibnamefont{Raussendorf}} \bibnamefont{and}
  \bibinfo{author}{\bibfnamefont{H.~J.} \bibnamefont{Briegel}},
  \bibinfo{journal}{Phys. Rev. Lett.} \textbf{\bibinfo{volume}{86}},
  \bibinfo{pages}{5188} (\bibinfo{year}{2001}).

\bibitem[{\citenamefont{Childs et~al.}(2004)\citenamefont{Childs, Leung, and
  Nielsen}}]{general_proj_meas}
\bibinfo{author}{\bibfnamefont{A.~M.} \bibnamefont{Childs}},
  \bibinfo{author}{\bibfnamefont{D.~W.} \bibnamefont{Leung}}, \bibnamefont{and}
  \bibinfo{author}{\bibfnamefont{M.~A.} \bibnamefont{Nielsen}}
  (\bibinfo{year}{2004}), \eprint{quant-ph/0404132}.

\bibitem[{\citenamefont{Bennett et~al.}(1993)\citenamefont{Bennett, Brassard,
  Crepeau, Jozsa, Peres, and Wootters}}]{teleportation}
\bibinfo{author}{\bibfnamefont{C.}~\bibnamefont{Bennett}},
  \bibinfo{author}{\bibfnamefont{G.}~\bibnamefont{Brassard}},
  \bibinfo{author}{\bibfnamefont{C.}~\bibnamefont{Crepeau}},
  \bibinfo{author}{\bibfnamefont{R.}~\bibnamefont{Jozsa}},
  \bibinfo{author}{\bibfnamefont{A.}~\bibnamefont{Peres}}, \bibnamefont{and}
  \bibinfo{author}{\bibfnamefont{W.}~\bibnamefont{Wootters}},
  \bibinfo{journal}{Phys. Rev. Lett.} \textbf{\bibinfo{volume}{70}},
  \bibinfo{pages}{1895} (\bibinfo{year}{1993}).

\bibitem[{\citenamefont{Fisher}(1998)}]{spin_liquid}
\bibinfo{author}{\bibfnamefont{M.~P.~A.} \bibnamefont{Fisher}}
  (\bibinfo{year}{1998}), \eprint{cond-mat/9806164}.

\bibitem[{\citenamefont{Popp et~al.}(2004)\citenamefont{Popp, Verstraete,
  Martin-Delgado, and Cirac}}]{localizable_entanglement}
\bibinfo{author}{\bibfnamefont{M.}~\bibnamefont{Popp}},
  \bibinfo{author}{\bibfnamefont{F.}~\bibnamefont{Verstraete}},
  \bibinfo{author}{\bibfnamefont{M.~A.} \bibnamefont{Martin-Delgado}},
  \bibnamefont{and} \bibinfo{author}{\bibfnamefont{J.~I.} \bibnamefont{Cirac}}
  (\bibinfo{year}{2004}), \eprint{quant-ph/0411123}.

\bibitem[{\citenamefont{Majumdar and Ghosh}(1969)}]{mg}
\bibinfo{author}{\bibfnamefont{C.~K.} \bibnamefont{Majumdar}} \bibnamefont{and}
  \bibinfo{author}{\bibfnamefont{D.~K.} \bibnamefont{Ghosh}},
  \bibinfo{journal}{J. Math. Phys.} \textbf{\bibinfo{volume}{10}},
  \bibinfo{pages}{1399} (\bibinfo{year}{1969}).

\bibitem[{\citenamefont{Chayes et~al.}(1989)\citenamefont{Chayes, Chayes, and
  Kivelson}}]{rvb_singlet}
\bibinfo{author}{\bibfnamefont{J.}~\bibnamefont{Chayes}},
  \bibinfo{author}{\bibfnamefont{L.}~\bibnamefont{Chayes}}, \bibnamefont{and}
  \bibinfo{author}{\bibfnamefont{S.}~\bibnamefont{Kivelson}},
  \bibinfo{journal}{Commun. Math. Phys.} \textbf{\bibinfo{volume}{123}},
  \bibinfo{pages}{53} (\bibinfo{year}{1989}).

\bibitem[{\citenamefont{Soos and Ramasesha}(1984)}]{soos_ramasesha}
\bibinfo{author}{\bibfnamefont{Z.~G.} \bibnamefont{Soos}} \bibnamefont{and}
  \bibinfo{author}{\bibfnamefont{S.}~\bibnamefont{Ramasesha}},
  \bibinfo{journal}{Phys. Rev. B} \textbf{\bibinfo{volume}{29}},
  \bibinfo{pages}{5410} (\bibinfo{year}{1984}).

\bibitem[{\citenamefont{Ising}(1925)}]{Ising}
\bibinfo{author}{\bibfnamefont{E.}~\bibnamefont{Ising}}, \bibinfo{journal}{Z.
  Phys.} \textbf{\bibinfo{volume}{31}}, \bibinfo{pages}{253}
  (\bibinfo{year}{1925}).

\bibitem[{\citenamefont{Affleck et~al.}(1987)\citenamefont{Affleck, Kennedy,
  Lieb, and Tasaki}}]{aklt}
\bibinfo{author}{\bibfnamefont{I.}~\bibnamefont{Affleck}},
  \bibinfo{author}{\bibfnamefont{T.}~\bibnamefont{Kennedy}},
  \bibinfo{author}{\bibfnamefont{E.~H.} \bibnamefont{Lieb}}, \bibnamefont{and}
  \bibinfo{author}{\bibfnamefont{H.}~\bibnamefont{Tasaki}},
  \bibinfo{journal}{Phys. Rev. Lett.} \textbf{\bibinfo{volume}{59}},
  \bibinfo{pages}{799} (\bibinfo{year}{1987}).

\bibitem[{\citenamefont{Verstraete et~al.}(2004)\citenamefont{Verstraete,
  Mart{\'i}n-Delgado, and Cirac}}]{div_ent_length}
\bibinfo{author}{\bibfnamefont{F.}~\bibnamefont{Verstraete}},
  \bibinfo{author}{\bibfnamefont{M.~A.} \bibnamefont{Mart{\'i}n-Delgado}},
  \bibnamefont{and} \bibinfo{author}{\bibfnamefont{J.~I.} \bibnamefont{Cirac}},
  \bibinfo{journal}{Phys. Rev. Lett.} \textbf{\bibinfo{volume}{92}}
  (\bibinfo{year}{2004}), \eprint{quant-ph/0311087}.

\bibitem[{\citenamefont{White}(1992)}]{dmrg}
\bibinfo{author}{\bibfnamefont{S.~R.} \bibnamefont{White}},
  \bibinfo{journal}{Phys. Rev. Lett.} \textbf{\bibinfo{volume}{69}},
  \bibinfo{pages}{2863} (\bibinfo{year}{1992}).

\bibitem[{\citenamefont{Hagiwara et~al.}(1990)\citenamefont{Hagiwara,
  Katsumata, Affleck, Halperin, and Renard}}]{spinhalf_ends}
\bibinfo{author}{\bibfnamefont{M.}~\bibnamefont{Hagiwara}},
  \bibinfo{author}{\bibfnamefont{K.}~\bibnamefont{Katsumata}},
  \bibinfo{author}{\bibfnamefont{I.}~\bibnamefont{Affleck}},
  \bibinfo{author}{\bibfnamefont{B.~I.} \bibnamefont{Halperin}},
  \bibnamefont{and} \bibinfo{author}{\bibfnamefont{J.~P.}
  \bibnamefont{Renard}}, \bibinfo{journal}{Phys. Rev. Lett.}
  \textbf{\bibinfo{volume}{65}}, \bibinfo{pages}{3181} (\bibinfo{year}{1990}).

\bibitem[{\citenamefont{Fan et~al.}(2004)\citenamefont{Fan, Korepin, and
  Roychowdhury}}]{fan_aklt}
\bibinfo{author}{\bibfnamefont{H.}~\bibnamefont{Fan}},
  \bibinfo{author}{\bibfnamefont{V.}~\bibnamefont{Korepin}}, \bibnamefont{and}
  \bibinfo{author}{\bibfnamefont{V.}~\bibnamefont{Roychowdhury}},
  \bibinfo{journal}{Phys. Rev. Lett.} \textbf{\bibinfo{volume}{93}}
  (\bibinfo{year}{2004}).

\bibitem[{\citenamefont{Ashketar and Lewandowski}(1994)}]{haar}
\bibinfo{author}{\bibfnamefont{A.}~\bibnamefont{Ashketar}} \bibnamefont{and}
  \bibinfo{author}{\bibfnamefont{J.}~\bibnamefont{Lewandowski}},
  \emph{\bibinfo{title}{Representation theory of analytic holonomy C*
  algebras}} (\bibinfo{publisher}{Oxford University Press},
  \bibinfo{address}{Oxford}, \bibinfo{year}{1994}).

\bibitem[{\citenamefont{den Nijs and Rommelse}(1989)}]{preroughening}
\bibinfo{author}{\bibfnamefont{M.}~\bibnamefont{den Nijs}} \bibnamefont{and}
  \bibinfo{author}{\bibfnamefont{K.}~\bibnamefont{Rommelse}},
  \bibinfo{journal}{Phys. Rev. B} \textbf{\bibinfo{volume}{1989}},
  \bibinfo{pages}{4709} (\bibinfo{year}{1989}).

\end{thebibliography}
\end{document}